# Distinguishing True Influence from Hyperprolificity with Citation Distance

Lu Li[a], Yun Wan[b], Feng Xiao[a,1]

[a] *Business School, Sichuan University, PR China*

[b] *School of Economics and Management, Anhui University of Science and Technology*

**Abstract:** Accurately evaluating scholarly influence is essential for fair academic assessment, yet traditional bibliometric indicators—dominated by publication and citation counts—often favor hyperprolific authors over those with deeper, long-term impact. We propose the $x$-index, a novel citation-based metric that conceptualizes citation as a process of knowledge diffusion and incorporates citation distance to reflect the structural reach of scholarly work. By weighting citations according to the collaborative proximity between citing and cited authors, the $x$-index captures both the depth and breadth of influence within evolving academic networks. Empirical analyses show that the $x$-index significantly improves the rankings of Turing Award recipients while reducing those of hyperprolific authors, better aligning rankings with recognized academic merit. It also demonstrates superior discriminatory power among early-career researchers and reveals stronger sensitivity to institutional research quality. These results suggest that the $x$-index offers a more equitable and forward-looking alternative to existing metrics, with practical applications in talent identification, funding decisions, and academic recommendation systems.

Over the past two decades, quantitative citation-based indicators have become deeply embedded in the evaluation and governance of modern science [1]. Metrics such as number of publications (*np*), total citations (*tc*), and the $h$-index [2] are widely used to assess researchers, allocate funding, and shape institutional rankings. Their appeal lies in standardization and scale: they offer simple, comparable proxies for scholarly influence across disciplines. However, growing evidence suggests that these indicators do not merely measure scientific activity; they also shape research behavior and career trajectories [3-5]. A widely cited study published in *Nature* reports that thousands of scientists produce, on average, one new paper every five days [6], raising concerns that hyperprolific authors—rather than field-shaping thinkers—are increasingly dominating the academic leaderboard. As science becomes more interdisciplinary and collaborative, the question becomes increasingly pressing: do existing metrics reward lasting

---

[1] Corresponding author. Email: evan.fxiao@gmail.com.

influence, or simply prolific visibility?

This tension becomes particularly evident in recent patterns of scientific production. Large-scale empirical analyses show that a growing number of researchers now publish at rates far exceeding historical norms [7, 8]. While such accelerated output may partly reflect large-team science, intensified collaboration, or evolving authorship practices, it also raises questions about how scholarly influence is signaled and interpreted. In highly collaborative environments, frequent publication can generate dense and localized citation activity [9], rapidly increasing visibility without necessarily implying broad or enduring intellectual reach. As a result, contributions whose ideas diffuse widely across scientific communities may be less prominently recognized than those associated with concentrated, high-volume output.

Widely used bibliometric indicators share a common limitation in how they operationalize scientific influence. Quantity-based metrics, such as $np$ and $tc$, treat scholarly output as additive and homogeneous, implicitly assuming that each publication and citation contributes equally to influence [10], regardless of where attention originates or how it propagates across the scientific community [11, 12]. The $h$-index [2] attempts to balance quantity and quality by discounting low-citation outputs, yet it still aggregates citations without considering structural differences in citation pathways or audience diversity [13]. Network-based measures, including PageRank-style metrics [14-16], incorporate relationships among authors, yet they remain inherently affected by network dynamics and the "rich-get-richer" effect [17, 18], whereby well-connected scholars receive disproportionate recognition. As a result, prevailing evaluation metrics tend to capture productivity and positional prominence more effectively than the structural reach of scientific ideas.

To examine how prevailing bibliometric indicators distinguish between productivity-driven visibility and scientific influence, we contrast two empirically and conceptually distinct groups of scholars: **hyperprolific authors (HPs)**—defined as those who published more than 72 papers in any single year during the observation period [6], capturing extreme levels of publication output—and **Turing Award recipients (TAs)**, who are widely acknowledged for foundational and field-shaping contributions in computer science, serving as an expert-validated benchmark of long-term scientific influence. Using a large-scale citation dataset comprising over 73 million citation relationships up to July 2025 [19], we identified 231 HPs and 59 TAs from 1980 to 2024, with **no overlap** between the two groups. This contrast provides a natural testbed for assessing whether commonly used metrics prioritize volume-driven prominence or structurally diffused scientific impact.

When scholars are ranked by conventional bibliometric indicators, the contrast between these two groups becomes pronounced. As shown in **Fig. 1b**, HPs are substantially more prevalent than TAs among the top 1,000 scholars ranked by *np*, *tc*, $h$-index, and pagerank score(*pr*). For example, ranking by the $h$-index identifies 142 HPs but only 13 TAs, with similarly skewed patterns observed across all examined metrics. When normalized by category size, the disparity remains pronounced: the $h$-index captures 142 of 231 HPs (61.5%) but only 13 of 59 TAs (22.03%). This complete absence of overlap, together with the overrepresentation of HPs among top-ranked scholars, suggests that traditional bibliometric indicators and expert-recognized scientific prestige capture largely orthogonal dimensions of scholarly influence.

This misalignment is further reinforced by temporal trends. **Fig. 1a** illustrates that HPs first appeared in 2005 and have increased rapidly over time, growing from 11 individuals in 2016 to an astonishing 197 in 2024. The rapid expansion of extreme productivity highlights a structural shift in scientific production rather than isolated anomalies.

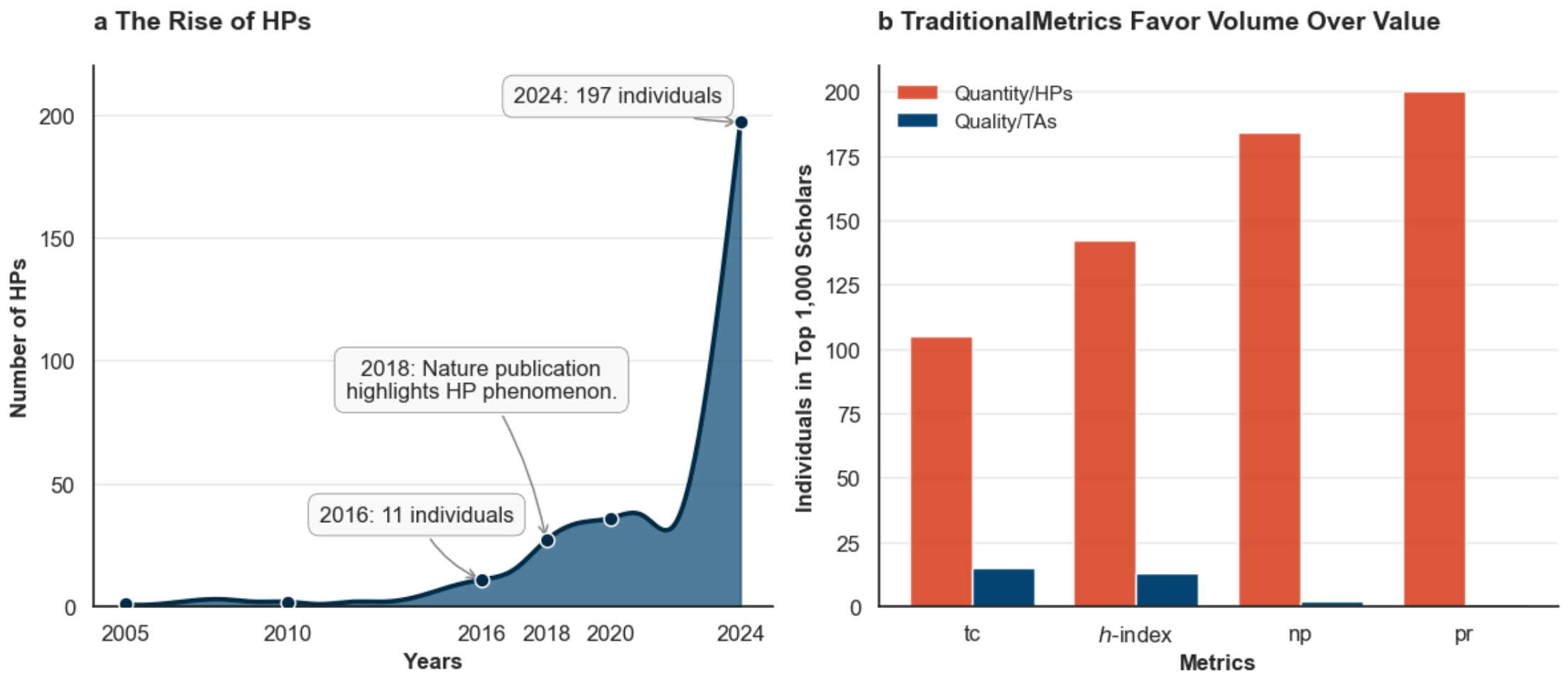

**Fig. 1 Temporal evolution of HPs and the inherent bias of traditional academic metrics. (a)** The explosive growth of HPs from 2005 to 2024. The solid line represents the number of individuals publishing at extreme rates, with key historical milestones highlighted, including the 2018 Nature study that brought international attention to the HP phenomenon. **(b)** Comparative representation of HPs and TAs within the top 1,000 globally ranked scholars across four traditional metrics in 2024: total citations (tc), *h*-index, number of publications (np), and pagerank score(pr). While traditional metrics heavily favor HPs (red bars), foundational thinkers like TAs (blue bars) are virtually invisible in the top echelons of these rankings. This disparity illustrates the low-resolution flaw of threshold-based metrics, which prioritize quantitative volume over field-shaping qualitative impact.

While extremely high publication rates can generate dense and localized citation activity, foundational contributions often exert influence in a more diffuse and long-range manner. Evidence from the contrast between HPs and TAs suggests that current evaluation metrics disproportionately reward concentrated productivity, while failing to capture long-term, systemic contributions that shape the discipline. These limitations suggest the need to reconsider how citations reflect scholarly influence.

Prior studies have therefore argued that citations should not be treated as isolated counts, but as signals of knowledge diffusion across scholarly communities[20-24]. From this perspective, a citation represents not merely recognition, but the transmission of ideas [25], with influence expanding as knowledge reaches more diverse and structurally distant audiences [26]. Citations can thus be understood as extensions of expert evaluation, where the "expert circle" grows beyond direct peers to encompass the broader scientific community. Under this diffusion-based view, impactful ideas are those that traverse structural boundaries—such as disciplines, institutions, or collaboration clusters—rather than simply accumulating citations from proximate networks [27]. This naturally motivates the use of citation distance, which captures the collaborative proximity between citing and cited authors, providing a principled way to quantify the structural reach of scientific contributions. Several prior studies have incorporated citation distance to account for long-range knowledge diffusion in research evaluation, among which the $c$-index [28] represents a notable step by explicitly modeling collaborative proximity between citing and cited authors. However, existing distance-aware indicators often rely on static networks, coarse distance thresholds, or restricted citation subsets, limiting their ability to capture the continuous and evolving nature of scientific influence. As collaborations evolve and knowledge flows grow increasingly complex, capturing both the depth and structural reach of scholarly influence requires a dynamic and interpretable evaluation framework.

Building on these insights, we introduce a structure-sensitive citation framework that models citations as diffusion events embedded in a dynamic collaboration network. Within this framework, we propose the $x$-index, a citation-based metric that aggregates citations by assigning each one a distance-dependent weight reflecting the collaborative proximity between citing and cited authors. Specifically, citations originating from structurally more distant collaborators contribute more strongly, while those arising within close collaboration neighborhoods receive relatively lower weight, through a continuous and temporally robust weighting scheme. In this way, the $x$-index integrates both the quantity of scholarly attention and the structural reach of influence, valuing not only how often a work is cited, but how far its influence travels across the academic landscape. Unlike classical metrics that treat all citations as homogeneous, or threshold-based distance measures that collapse citation distributions, or network centrality indicators that primarily emphasize a scholar's topological position [29], the $x$-index captures the cumulative diffusion-weighted influence of a scholar's work. It preserves fine-grained information about how ideas propagate across the scientific system, while remaining mathematically interpretable, temporally robust, and adaptable to dynamic collaboration networks.

Using a large-scale citation dataset, we show that the proposed framework systematically mitigates biases inherent in conventional evaluation metrics by distinguishing extreme productivity from structurally diffused scientific influence. The $x$-index enables more faithful identification of foundational contributors and emerging high-potential researchers across career stages. In the sections that follow, we formally define the $x$-index, examine its theoretical properties, and validate its performance through extensive empirical comparisons with existing indicators.

## Results

### Definitions

**Collaboration network and citation distance.** We construct a yearly co-authorship network $G_t = (V_t, E_t)$, where nodes represent authors, and an undirected unweighted edge $(i, j) \in E_t$ is added if $i$ and $j$ have coauthored at least one publication within a sliding window of the past five years $[t-4, t]$. This temporal window captures the empirically observed decay of collaboration ties while preserving a well-connected backbone of the field; compared with shorter windows that fragment the network and longer windows that retain stale ties and compress distances, a five-year horizon—widely adopted in bibliometric studies [30, 31]—offers a practical balance between recency and network stability.

For any citing–cited paper pair $(p, q)$ observed in year $t$, let $A_p$ and $A_q$ denote their respective sets of authors. We define the citation distance between $p$ and $q$ as the shortest co-authorship distance between the two author sets:

$$d_t(p, q) = \min_{i \in A_p, j \in A_q} dist_t(i, j)$$

where $dist_t(i, j)$ is the length of the shortest path between $i$ and $j$ in $G_t$, measured in hops. Distances are finite if both authors are in the same connected component of $G_t$, and set to infinity otherwise. A distance of zero is assigned if the two papers share at least one author. For notational simplicity, we omit the explicit dependence on $t$ in subsequent equations.

***x*-index**. Building on this framework, the $x$-index is defined as the cumulative sum of citations weighted by their structural distances in the collaboration network. The weight assigned to each citation quantifies the relative extent of its diffusion through the collaboration network.

$$w(d) = 1 - e^{-\frac{d}{\bar{d}}}$$

where $d$ is the citation distance and $\bar{d}$ is the normalization factor, representing the average citation distance in year $t$, excluding infinite values. This normalization facilitates comparability across time. We adopt the standard convention $\exp(-\infty) = 0$, which implies $w(\infty) = 1$, and $w(0) = 0$. The $x$-index of a scholar is then given by:

$$x = \sum_{i=1}^{N} w(d_i)$$

where $N$ is the total number of citations received by the scholar, and $d_i$ and $w(d_i)$ are the distance and weight of the $i$-th citation, respectively.

**Mathematical properties of the weighting function.**

We analyze the key mathematical properties of the exponential weighting function used in the x-index, including monotonicity, boundedness, and concavity.

**Property 1: Monotonicity**

**Statement.** If $\bar{d} > 0$, then $w(d)$is strictly increasing for all finite $d \geq 0$, and non-decreasing on $[0, \infty)]$.

**Proof**. For $d \in [0, \infty)$, the first derivative satisfies

$$\frac{\mathrm{d}w(d)}{\mathrm{d}d} = \frac{1}{\bar{d}} e^{-\frac{d}{\bar{d}}}$$

Since $d > 0$ and $e^{-\frac{d}{\bar{d}}} > 0$ for all $d \geq 0$, the first derivative is strictly positive (as illustrated in **Fig. 2b**). Together with the boundary convention $w(\infty) = \lim_{d \to \infty} w(d) = 1$, the function is non-decreasing on $[0, \infty)$.

**Property 2: Boundedness**

**Statement.** If $\bar{d} > 0$, then $w(d) \in [0,1]$ for all $d \in [0, \infty]$.

**Proof**. By Proposition 1, $w(d)$ is monotone increasing in $d$, so its minimum over $[0, \infty)$ occurs at $d = 0$ and its supremum occurs as $d \to \infty$. Evaluating the endpoints:

At $d = 0$, $e^{-\frac{d}{\bar{d}}} = e^0 = 1$, thus $w(0) = 1 - 1 = 0$.

As $d \to \infty$, $e^{-\frac{d}{\bar{d}}} \to 0$, thus $w(\infty) = 1 - 0 = 1$.

Therefore, for all $d \in [0, \infty]$, $0 \leq w(d) \leq 1$, i.e., $w(d)$ is bounded in $[0,1]$ (as demonstrated in **Fig. 2a**).

**Property 3: Concavity**

**Statement.** If $\bar{d} > 0$, then $w(d)$ is concave on $[0, \infty)$.

**Proof**. For $d \in [0, \infty)$, the second derivative satisfies

$$\frac{\mathrm{d}^2 w(d)}{\mathrm{d}d^2} = -\frac{1}{\bar{d}^2} e^{-\frac{d}{\bar{d}}}$$

Because $\bar{d} > 0$ and $e^{-\frac{d}{\bar{d}}} > 0$, we have $\frac{d^2 w}{dd^2} < 0$. Hence $w(d)$ is strictly concave on $[0, \infty)$ (as demonstrated in **Fig. 2c**). Concavity means the incremental increase in weight decreases as $d$ grows, capturing diminishing returns for increasingly distant citations.

**Practical note on feasibility of $\bar{d} > 0$**

The condition $\bar{d} > 0$ holds whenever at least one finite citation distance with positive length exists in year $t$. Since $\bar{d}$ is computed from finite distances (excluding non-finite values), it is well-defined for years in which the collaboration network contains at least one reachable citing–cited author pair.

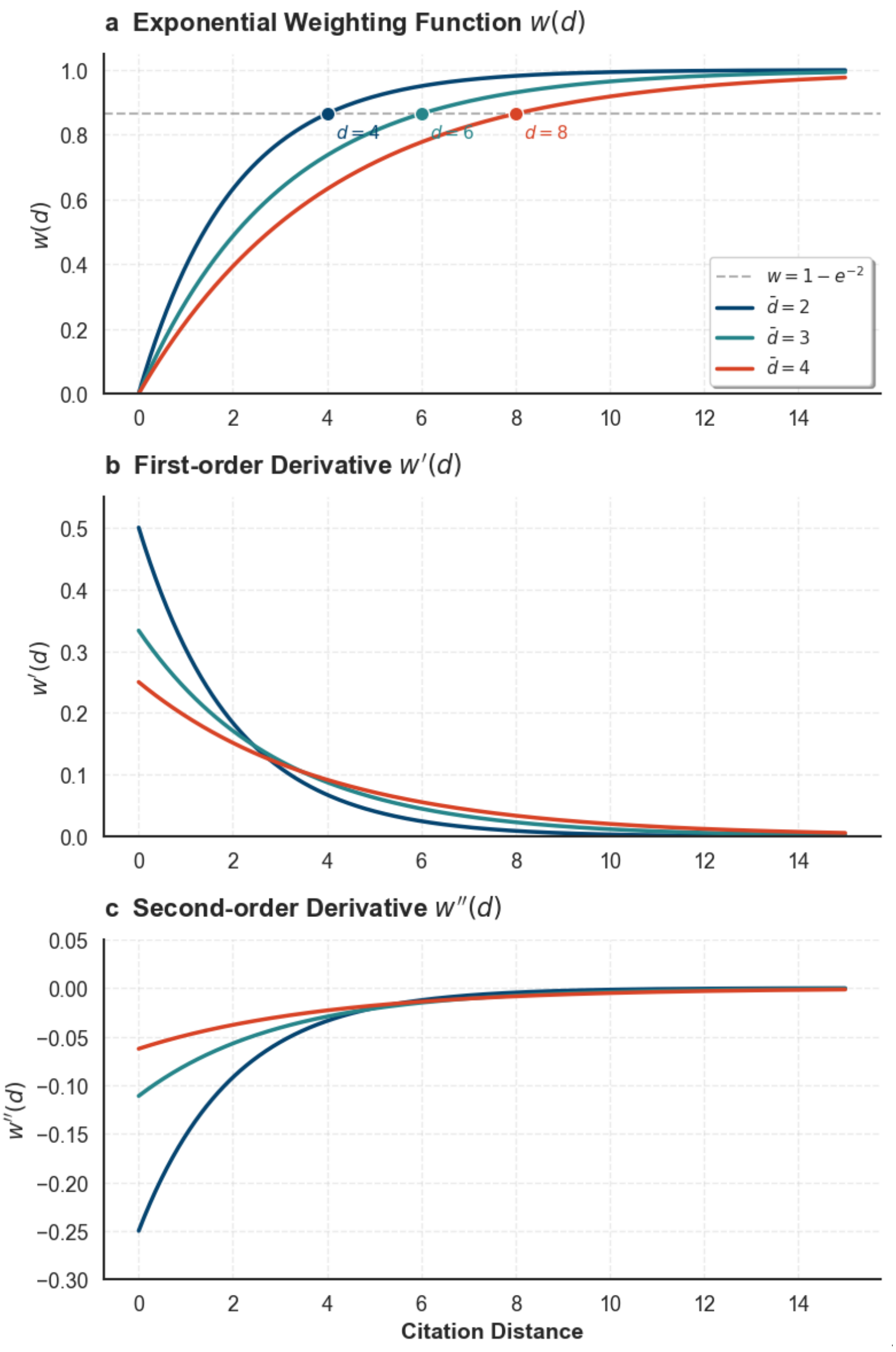

**Fig. 2 Mathematical properties and scalability of the exponential weighting function**. **(a)** The exponential weighting function $w = 1 - e^{-d/\bar{d}}$, normalized within the unit interval $[0, 1]$. Solid curves represent the function under varying mean finite citation distances $\bar{d} \in \{2,3,4\}$. The horizontal dashed line identifies the reference value $w = 1 - e^{-2} \approx 0.865$; markers indicate the specific points where $d = 2\bar{d}$, demonstrating that the weighting remains invariant relative to the scale parameter $\bar{d}$. This normalization ensures that evaluation metrics remain robust and consistent across evolving scientific network topologies. **(b)** The first-order derivative of $w'(d)$, which is strictly positive for all $d \geq 0$, ensuring the function's strict monotonicity in penalizing citation distance. **(c)** The second-order derivative of $w''(d)$, which remains strictly negative for all $d \geq 0$, confirming the concave nature of the weighting growth and its diminishing marginal sensitivity as distance increases.

## Citation-distance metrics attenuate productivity-driven ranking differences

To examine whether citation-distance metrics mitigate the well-known productivity bias in traditional bibliometric indicators, we compared ranking behaviors across two contrasting groups of scholars: TAs and HPs. Using the Aminer DBLP-Citation-network V18 dataset [19], we identified 59 TAs and 231 HPs between 1980 and 2024. For each author $a$, we computed the ranking shift induced by the $x$-index relative to a baseline metric as $r_a = r_a^x - r_a^{\text{baseline}}$, where $r_a^x$ and $r_a^{\text{baseline}}$ represent the ranking of author $a$ under the $x$-index and the baseline metric, respectively. A negative $\Delta r_a$ indicates a ranking improvement under the $x$-index relative to the baseline metric.

For TAs, we evaluated rankings in the award year to avoid the inflationary effect of post-award citations. A one-sided Wilcoxon signed-rank test [32] was used to evaluate the hypothesis that the median ranking shift is negative. Across all baseline metrics—$np$, $tc$, $h$-index, and $pr$—the test consistently rejected the null hypothesis (**Fig. 3a**, all $p < 0.05$), indicating that the $x$-index assigns TAs significantly better rankings than traditional measures. For HPs, we used rankings as of 2024 and tested the reverse hypothesis-the median ranking shift is positive. In contrast to TAs, the $x$-index produced significantly lower rankings for HPs compared with $np$, $tc$, $h$-index, and $pr$ (all $p < 0.05$), indicating that authors who achieve high visibility primarily through volume tend to be down-ranked once structural diffusion is taken into account. Additional statistical tests of $c$-index yield similar pattern for HPs (**Fig. 3b**, all $p < 0.05$), yet shows reduced statistical sensitivity in elevating TAs under the $tc$ baseline ($p > 0.05$). These results suggest that citation-distance metrics, particularly the $x$-index, systematically elevate scholars with deeply diffused and enduring impact while suppressing those whose prominence arises mainly from sheer productivity.

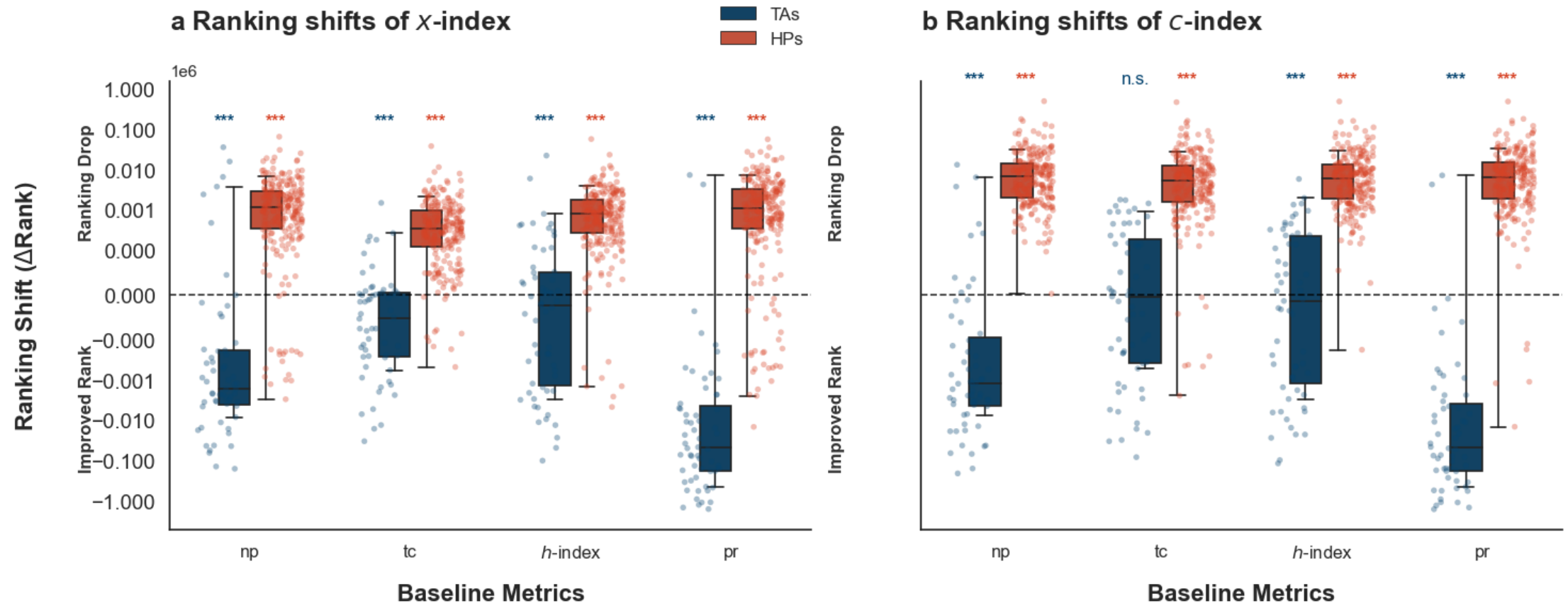

**Fig. 3 Comparative analysis of ranking shifts across citation-distance metrics**. **a-b,** Visualization of ranking shifts induced by the $x$-index (**a**) and the $c$-index (**b**) relative to four baseline metrics. Data points represent individual TA ($n = 59$, dark blue) and HP ($n = 231$, crimson). Ranking shift (Δrank) is calculated as $r_a^{metric} - r_a^{\text{baseline}}$, where negative values indicate improved rankings and positive values represent a ranking drop. Boxplots show the median and interquartile range (IQR), while jittered points reveal the underlying distribution. All statistical tests were one-sided Wilcoxon signed-rank test. Asterisks denote significance levels: $*** \ p < 0.001$; $** \ p < 0.01$; $* \ p < 0.05$, the abbreviation **n.s.** (not significant) denotes a $p > 0.05$. **(a)** The $x$-

index consistently elevates the rankings of TAs while significantly suppressing HPs across all metrics ($p < 0.05$), demonstrating a robust correction for productivity bias. **(b)** The $c$-index exhibits a qualitatively similar pattern for HPs yet shows reduced statistical sensitivity in elevating TAs under the $tc$ baseline ($p > 0.05$). These results suggest that citation-distance metrics, particularly the $x$-index, systematically elevate scholars with deeply diffused and enduring impact while suppressing those whose prominence arises mainly from sheer productivity.

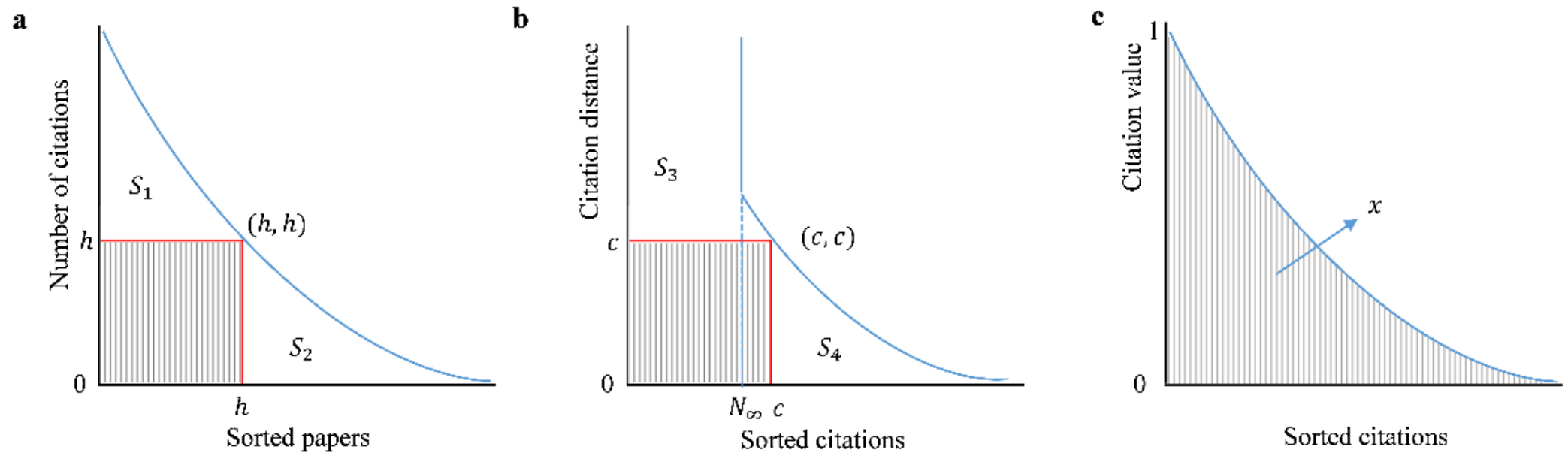

**Fig. 4 Comparative visualization of the $h$-index, $c$-index and $x$-index. (a)** Citation distribution curve, where $tc$ as the area under the curve. The $h$-index corresponds to the largest square beneath the curve, with one vertex at the origin $(0,0)$. Shaded regions $S_1$ and $S_2$ denote contributions from highly cited and low-cited papers, respectively. **(b)** The $c$-index identifies the largest square under the citation distribution curve with one vertex at the origin $(0,0)$, with regions $S_3$ and $S_4$ representing short-distance and long-distance citations, respectively. $N_\infty$ denotes the count of citations originating from completely disconnected network components. **(c)** The $x$-index aggregates all citations weighted by citation distance, transformed through a value function that accounts for citation distance and network connectivity.

## Why $x$-index outperforms $c$-index: mechanism and robustness analysis

Although the $x$-index and the $c$-index both leverage citation distance and exhibited similar directional effects in distinguishing TAs from HPs, their behaviors diverge once the structural properties of citation-distance distributions are examined in detail. As shown in **Fig. 4b**, citations are ranked by structural distance in descending order. The $c$-index identifies the largest square $(c,c)$ under the citation distance curve. However, it captures only the top $c$ citations and disregards contributions from long-distance citations beyond the threshold ($S_3$) as well as short-distance citations below the threshold ($S_4$). This selective evaluation leads to three structural limitations that become pronounced at different stages of a scholar's career and under different network conditions.

**Limitation 1: Information loss in the small $c$-index regime (early-career scholars).** The first limitation is most pronounced early in a scholar's career, when the $c$-index is typically small and therefore acts as a coarse thresholding measure of diffusion breadth. When $c$ is small, citations with distances greater than or equal to $c$ (**Fig. 4** region $S_3$) often contain substantial structural information, yet the $c$-index records only whether this region contains at least $c$citations. Consequently, scholars with markedly different diffusion breadths may receive identical $c$-index values. This effect is illustrated by scholars whose $c$-index in 2024 equals 5. As shown in **Fig. 5a**, this group comprises 344,861 scholars whose $x$-index values range from 3 to 953, spanning more than two orders of magnitude. **Fig. 5b** contrasts two representative cases: **Scholar 1** has five citations, all exactly at distance 5, whereas **Scholar 2** has twenty-two citations, eighteen of which fall in $S_3$ region ($d \geq 5$). Despite these stark differences in the size and shape of their finite-distance diffusion, the $c$-index records both scholars as identical. In other words, when $c$ is small, the entire structure of $S_3$—the main body of meaningful finite-distance citations—is elided by the $c$-index.

**A second limitation emerges in later career stages, when infinite or near-infinite citation distances become more common**. Once the $c$-index exceeds the maximum finite distance attainable in the network, its value becomes determined almost entirely by the number of infinite-distance citations. To evaluate the magnitude of this information loss, we analyzed all scholars whose $c$-index in 2024 equals 25. As shown in **Fig. 5a**, this group includes 5,077 scholars whose $x$-index values range from 25 to 2,85—a spread exceeding two orders of magnitude. **Fig. 5c** illustrates two representative scholars: **Scholar 3** had only twenty-five citations, all contributing to the $c$-index, whereas **Scholar 4** had 455 citations, 94.51% of which had distances below 25 and were therefore ignored.

**Limitation 3: Temporal instability under evolving network topology.** The third and most fundamental limitation stems from the dynamic evolution of the global collaboration network. As shown in **Fig. 6b**, the share of infinite citation distances declines sharply from nearly 80% in the 1950s to below 10% in recent years, while the distribution of finite distances exhibits a rise-then-decline pattern, with median distances decreasing steadily over the past three decades in **Fig. 6a**. These trends reflect a structural contraction of the collaboration graph as scientific communities become more interconnected.

Under such shifting topologies, the absolute meaning of a fixed citation distance changes dramatically over time: For example, in 1980, ten citations at distance 5 represented exceptional structural reach, whereas in 2000 the same configuration represented a relatively common diffusion pattern, as collaboration paths were significantly shorter. Because the $c$-index applies a fixed threshold to raw distances, it treats these scenarios identically, even as those distances shrink or expand with the evolving network, rendering its values sensitive to global network evolution and temporally incomparable.

Together, these three limitations—information loss in the long-distance region, insensitivity to the magnitude and composition of the finite-distance tail, and vulnerability to global network evolution—highlight the drawbacks of threshold-based citation-distance metrics. By contrast, the $x$-index evaluates every citation through a continuous, saturating weight function normalized by the contemporaneous mean finite distance. This design preserves the relative value of long-distance citations across different historical periods, captures the full shape of diffusion in both $S_3$ and $S_4$, and avoids the abrupt phase transitions inherent to threshold-based metrics. Consistent with these mechanisms, the $x$-index exhibits the highest coefficient of variation among all evaluated metrics (**Table 1**), indicating superior discriminative resolution across scholars. As a result, the $x$-index provides a more complete, robust, and temporally consistent characterization of structural diffusion.

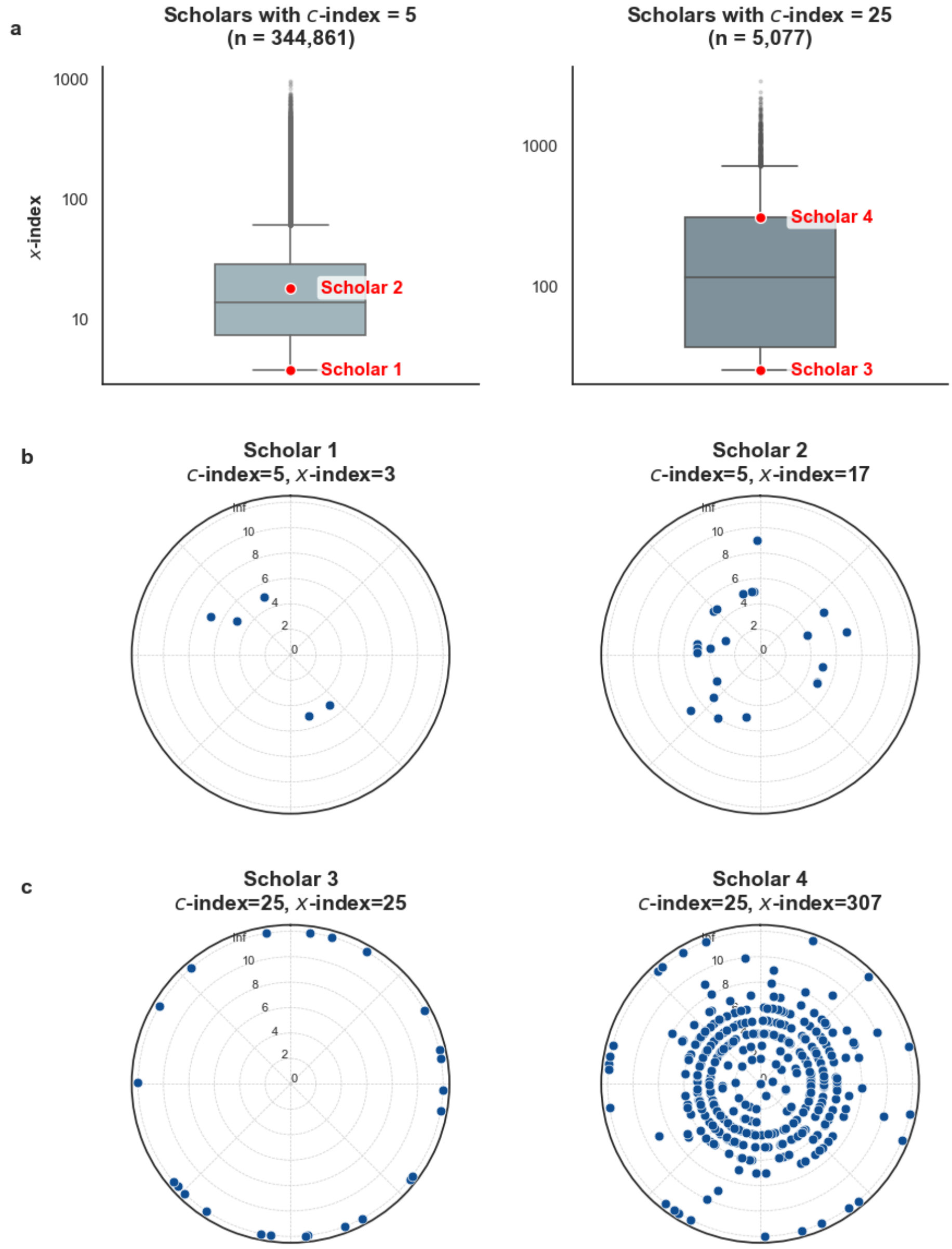

**Fig. 5 Macro-distribution and micro-evidence of scholars' $x$-index under fixed $c$-index constraints. (a)** Distribution of $x$-index for scholars with a fixed $c$-index of 5 ($n = 344{,}861$) and 25 ($n = 5{,}077$). Boxplots indicate the median, quartiles, and range of the $x$-index within each group. Red dots highlight four representative scholars (Scholars 1–4) selected for micro-level citation analysis. **(b)** Comparison between Scholar 1 ($x$-index= 3) and Scholar 2 ($x$-index= 17) at $c = 5$. **(c)** Comparison between Scholar 3 ($x$-index= 25) and Scholar 4 ($x$-index= 307) at $c = 25$. Each blue dot represents a citation, where the radial position denotes the absolute citation distance $d$. Citations with infinite distance are normalized to the boundary ring at $d = 12$. Since the maximum measurable finite distance of these 4 scholars is $d = 10$, this intentional overflow allows

for the clear visualization of citations that exceed the local structural reach. Values shown are integer-valued $x$-index $\lfloor x \rfloor$ for readability; statistical tests are performed on the underlying continuous $x$.

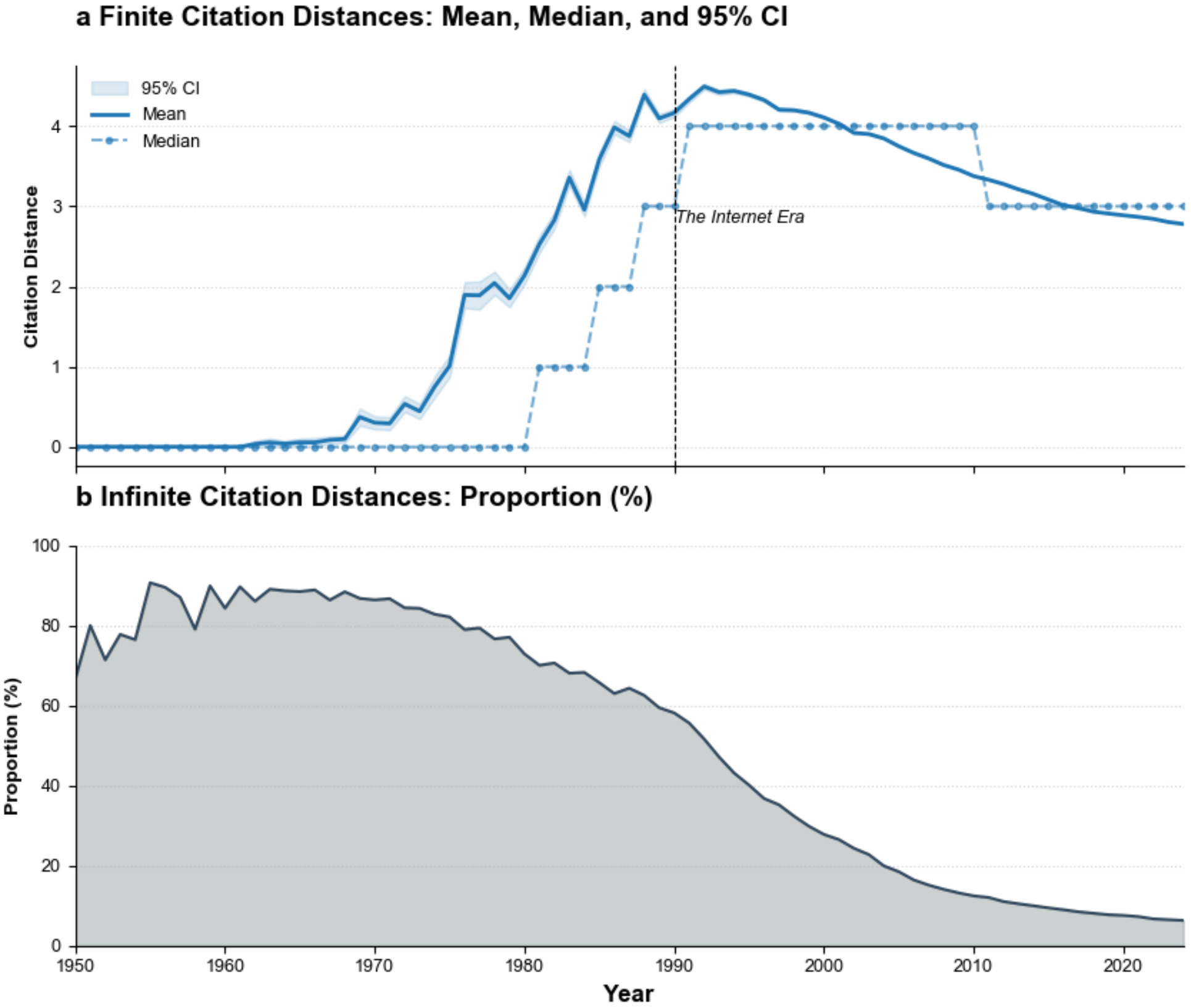

**Fig. 6 Temporal dynamics of citation connectivity and distances (1950–2024)**. The chart illustrates the evolution of scientific network structure across seven decades. **(a)** Trends in finite citation distances, where the solid blue line represents the annual mean and the dashed line indicate the annual median. The light blue shaded area denotes the 95% confidence interval (CI), reflecting the statistical stability of distance estimates. **(b)** The proportion of infinite citation distances over time, representing the degree of network fragmentation. The vertical dashed line marks the onset of The Internet Era, highlighting a significant structural shift toward a more integrated, Small World global scientific collaboration network.

**Table 1 Coefficient of Variation of metrics for all scholars in 2024**

| index | CV |
| --- | --- |
| $np$ | 3.131 |
| $tc$ | 9.169 |
| $h$-index | 1.703 |
| $pr$ | 1.859 |
| $c$-index | 7.060 |
| **$x$-index** | **9.330** |

## Institutional-level analysis

To extend the individual-level findings to a broader structural context, we next examine **whether similar diffusion patterns emerge at the institutional level**. Using the QS 2025 university rankings, we categorized scholars into two groups: T1, comprising authors affiliated with top 200 universities, and T3, consisting of those affiliated with universities ranked below 600 or unlisted. We identified scholars who were active during 2020–2024 and had accumulated at least 100 total citations by the end of 2024, yielding a working sample of 164,310 individuals. Primary affiliation was assigned based on the most recent institutional record, resulting in 111,172 scholars (67.7%) with available affiliation information (26,574 in T1 and 59,629 in T3). This classification parallels the structural roles observed in earlier analyses: T1 institutions typically provide visibility, international connectivity, and access to central collaboration structures analogous to those enjoyed by TAs, whereas T3 institutions reflect the more localized collaboration environments characteristic of ordinary or HPs. This tier-based perspective allows us to test whether distance-based citation metrics, particularly the $x$-index, can detect structural disparities at the institutional level, and whether the limitations observed for the $c$-index persist under institutional aggregation.

A direct comparison of metric distributions across T1 and T3 institutions reveals a striking divergence between the $x$-index and the $c$-index. Here we assess whether the *absolute metric values* of the two groups originate from different underlying distributions. Using one-sided Mann–Whitney U tests tailored to this objective, we find that scholars at T1 institutions exhibit substantially higher $x$-index values than those at T3 institutions ($p < 0.05$; **Table 2**), indicating broader and more structurally dispersed diffusion of their work. In contrast, the $c$-index shows no statistically significant difference between T1 and T3 ($p > 0.05$), implying that threshold-based distance metrics fail to detect the substantial structural disparities present across institutional tiers.

Further analysis shows that productivity-oriented metrics similarly fail to differentiate institutional tiers. As summarized in **Table 2**, neither $np$ nor $h$-index exhibit significant differences between T1 and T3 scholars ($p > 0.05$). This suggests that scholars at tail institutions can achieve publication volume and citation thresholds comparable to those at elite institutions, potentially as **a strategy** to counterbalance structural disadvantages in collaboration opportunities and external visibility.

These findings highlight that institutional differences in scholarly influence arise not from disparities in publication volume but from differences in the quality and structural reach of research output. Scholars at elite institutions tend to produce work that propagates more broadly across the collaboration network, yielding consistently higher $x$-index values. By weighting citations according to their structural distances, the $x$-index captures the extent to which a scholar's contributions travel beyond their immediate research circle—an aspect fundamentally linked to the intrinsic quality, novelty, and generality of the work. In contrast, quantity-oriented metrics such as $np$, $h$, and the threshold-based $c$-index cannot distinguish such qualitative differences. Thus, the $x$-index provides a more faithful measure of genuinely influential scholarship, revealing patterns of diffusion that traditional indicators—and the $c$-

index—fail to detect.

**Table 2 Mann – Whitney U tests comparing metric distributions between T1 and T3 institutions.**

| index | T1 vs. T3 | |
|---|---|---|
| | statistic | p-value |
| $np$ | 7.229E+08 | 1.000 |
| $tc$ | 8.571E+08 | **0.000** |
| $h$-index | 7.914E+08 | 0.601 |
| $pr$ | 7.397E+08 | 1.000 |
| $c$-index | 7.288E+08 | 1.000 |
| $x$-index | 8.408E+08 | **0.000** |

## Discussion

Scientific influence is commonly evaluated through productivity- and count-based indicators that implicitly equate impact with volume or visibility. This study introduces the $x$-index, a citation-distance-weighted metric that reframes scholarly influence as a process of diffusion through the collaboration network. By explicitly accounting for how far scientific ideas propagate across collaborative structures, the $x$-index captures both the scale and the structural breadth of scientific influence that is largely obscured by traditional bibliometric measures. In doing so, our work moves beyond productivity-oriented evaluation toward a diffusion-based perspective on scientific impact.

Our analyses demonstrate that structural diffusion constitutes a dimension of influence that is systematically overlooked by traditional indicators. While productivity-based metrics and threshold-based distance measures can reward volume or localized prominence, they compress or discard information about how influence spreads across communities. By preserving the full distribution of citation distances through a continuous weighting scheme, the $x$-index retains fine-grained structural information and distinguishes between concentrated visibility and broadly diffused intellectual reach. This distinction becomes evident not only at the individual level—where the $x$-index separates TAs from HPs—but also at the institutional level, where it reveals disparities in diffusion patterns that conventional metrics fail to detect.

These findings suggest that differences in scholarly influence across individuals and institutions arise less from sheer publication output than from variation in the structural reach of knowledge dissemination. By quantifying how contributions travel beyond immediate collaboration neighborhoods, the $x$-index provides a complementary lens for research evaluation, particularly in contexts where identifying durable, field-shaping influence is essential. Rather than replacing existing indicators, the $x$-index augments current evaluation frameworks with structurally informed evidence that is orthogonal to productivity and local visibility.

Several limitations merit consideration. Our analyses are based on citation data from English-language computer science publications, and future work should assess the $x$-index's generalizability across disciplines and multilingual contexts. In addition, extending the $x$-index to dynamic knowledge graphs guided by large language models (LLMs) may further enhance its applicability, enabling real-time influence estimation and context-aware citation analysis. Despite these limitations, the $x$-index advances a principled shift toward structural perspectives on scientific impact, emphasizing that influence is defined not only by how much attention a work receives, but also by how far that attention travels through the scientific system.

## Methods

**Data.** We use the DBLP-Citation-network V18 dataset from the Aminer Citation Network Dataset [19], a large-scale, disambiguated citation network covering the computer science literature. The dataset integrates records from multiple sources, including DBLP, ACM, and MAG (Microsoft Academic Graph), and provides detailed metadata for each publication, including abstracts, authors, publication years, venues, titles, and references. This dataset contains 6,733,427 papers and 73,702,758 citation links spanning more than 200 countries and territories up to July 2025. We retained only records with valid paper identifiers, publication years, and disambiguated author IDs, resulting in an analysis set of 6,532,072 papers, 71,721,509 citation links, and 3,457,761 unique scholars. TAs were identified based on the official listing published by the Association for Computing Machinery (ACM).

**Computation of the average citation distance.** Let $C_t$ be the set of citing–cited paper pairs observed in year $t$, and $C_t^{\mathrm{fin}} = \{(p,q) \in C_t : d(p,q) < \infty\}$. Then

$$\bar{d} = \frac{1}{C_t^{\mathrm{fin}}} \sum_{(p,q)\in C_t^{\mathrm{fin}}} d(p,q)$$

where $d(p,q)$ is the citation distance of the citing–cited pair $(p,q)$.

**Statistical analysis.** Data were processed using Python 3.9. Given the heavy-tailed nature of citation distributions, non-parametric tests were employed for all comparisons.

Matched Pair Analysis (TAs vs. HPs): To evaluate ranking shifts **Fig. 3**, we utilized the one-sided Wilcoxon signed-rank test for paired samples. The null hypothesis posited that the median ranking shift is non-negative for TAs (expecting improvement) or non-positive for HPs (expecting decline).

Institutional Analysis (T1 vs. T3): To compare metric distributions across independent institutional groups (**Table 2**), we performed the one-sided Mann–Whitney U test. Statistical significance was defined as $p < 0.05$.

**Baselines.** For comparison with existing bibliometric measures, we also compute several widely used indicators:

$np$: total number of papers authored by the scholar.

$tc$: total number of citations received by all publications authored by the scholar.

$h$-index: the largest integer $h$ such that the scholar has at least $h$ papers with $\geq h$ citations each.

$pr$: pagerank score of the scholar based on the collaboration network.

$c$-index: the largest integer $c$ such that the scholar has at least $c$ citations with distance greater than or equal to $c$.

All baseline metrics are computed using the same underlying publication and citation records as the x-index to ensure comparability.

## Acknowledgement

The work described in this paper was supported by the National Natural Science Foundation of China (72371174, 72531008, 72025104), the Natural Science Foundation of Sichuan Province (25NSFTD0032), and the Scientific Research Foundation for High-level Talents of Anhui University of Science and Technology (2024yjrc183).